\documentclass[12pt]{article}

\usepackage{graphicx}
\usepackage{subfigure}
\usepackage{amsmath}
\usepackage{amsfonts}
\usepackage{amssymb}
\usepackage{calligra}
\usepackage{bbold}

\def\be{\begin{equation}}
\def\ee{\end{equation}}
\def\bga{\begin{gather}}
\def\ega{\end{gather}}

\def\lsim{\mathrel{\rlap{\lower3pt\hbox{\hskip0pt$\sim$}}
     \raise1pt\hbox{$<$}}}         
\def\gsim{\mathrel{\rlap{\lower3pt\hbox{\hskip0pt$\sim$}}
     \raise1pt\hbox{$>$}}}         

\newcommand{\comment}[1]{}

\begin{document}

\numberwithin{equation}{section}

\begin{flushright}

{NYU-TH-01/26/2016}
\end{flushright}

\vskip 1cm

\begin{center}

{\Large {Nonlocal Galileons and Self-Acceleration}}

\end{center}

\thispagestyle{empty}

\vskip 0.5 cm

\begin{center}

{\large  {Gregory Gabadadze and Siqing Yu}}

 \vspace{.2cm}

{\it Center for Cosmology and Particle Physics,
Department of Physics,}
\centerline{\it New York University, New York,
NY, 10003, USA}

\end{center}

\vskip 1cm

\begin{abstract}
A certain class of nonlocal theories  eliminates an arbitrary cosmological constant (CC)
from a universe that can be perceived as our world. Dark energy then cannot be explained by 
a CC; it could however be due to massive gravity.  We calculate the 
new corrections, which originate from the nonlocal terms that eliminate the CC, to the decoupling limit 
Lagrangian of massive gravity. The new nonlocal terms 
also have internal field space Galilean symmetry and are referred here as ``nonlocal Galileons." 
We then study a self-accelerated solution and show that the new nonlocal 
terms change the perturbative stability analysis. In particular, small fluctuations are now stable and non-superluminal 
for some simple parameter choices, whereas for the same choices the pure massive gravity 
fluctuations are unstable.  We also study stable spherically symmetric solutions on this 
background.

\end{abstract}

\newpage

\section{Introduction and Summary}\label{intro}

Based on the ideas introduced in Refs. \cite{Linde:1988ws,Tseytlin:1990hn},   
the works  \cite{Gabadadze:2014rwa, ggsy}  proposed a class of theories that 
accomplish the following:  (1) they eliminate an arbitrary cosmological constant (CC) from our universe,   
(2) the  mechanism that eliminates CC is  stable with respect to the quantum loop corrections in particle 
physics and gravity.  The  feature  (1)  had been  achieved  in the original models of  
\cite{Linde:1988ws, Tseytlin:1990hn},  but (2) is  attained only in 
\cite{Gabadadze:2014rwa, ggsy}. A high price for the above virtues is  an intrinsic spacetime nonlocality  of these theories. However, this  nonlocality 
is sensed only by a CC, but not by any other form of matter or energy \cite{Gabadadze:2014rwa, ggsy}  (for earlier works addressing some of these issues, see \cite{ArkaniHamed:2002fu, Davidson}).

The present  work discusses the fate of  dark energy in the 
context of these CC-free theories: if a CC is removed entirely from the observed universe, then what is the substance  
that  leads to the accelerated expansion? The present cosmological data favors a ``dark fluid" with the equation of state $p=- \rho$.  Refs.   \cite{Gabadadze:2014rwa,ggsy}
propose to use massive gravity   \cite{mgr1,mgr2}, 
which gives rise to dark energy that is not a CC,  but has the equation of state $p=-\rho$  
\cite {dRGHP_sa,Theo,Koyama}  (or to use extensions of massive gravity \cite {extensions}, some of which are also 
known to have  healthy self-accelerated solutions, see Refs.  \cite {Mukohyama,Tanaka} and references therein).

Our focus will be on massive gravity 
\cite{mgr1,mgr2} and  its 
self-accelerated solution \cite {dRGHP_sa,Theo, Koyama},  
embedded in the framework of \cite {Gabadadze:2014rwa,ggsy}.  In particular, we will discuss 
this framework in the decoupling limit, where calculations 
are simple and transparent. This limit corresponds 
to the approximation that all scales considered are 
smaller than the Hubble scale. 

The requirement that the fluctuations of the self-accelerated solution contain no ghost constrains the 
parameter space of the theory. In pure massive gravity,  
the two free parameters, $a_2$ and $a_3$, 
have to be both nonzero and satisfy certain additional constraints to achieve this \cite {dRGHP_sa}. 
In particular, $a_3$ has to be nonzero and positive. 
For $a_3\neq 0$, the scalar-tensor Lagrangian obtained in the 
decoupling limit of massive gravity cannot be diagonalized 
due to a certain nonlinear term proportional to $a_3$. While this by itself  is not  a problem 
but only a technical issue, once we embed the pure massive gravity into the framework of 
\cite {ggsy}, we find that $a_3=0$ is a perfectly good choice 
for which the self-accelerated solution could be stable, given appropriate values of $a_2$.

More unsettling  is the  issue of vector fluctuations on the 
self-accelerated background  in pure massive gravity, who lose both their  kinetic  and the
gradient quadratic terms.  In a classical theory, the nonlinear 
terms are at least quadratic in the vector field, hence a trivial solution ``vector $=0$" could be enforced by the initial data.   
However, quantum fluctuations would invalidate these classical arguments: if no vector field kinetic term is induced by the loops, then the vector quantum fluctuations would be infinitely strongly coupled.
Alternatively,  the loops may generate  a kinetic term 
for the vector field,  and if it has a right sign, this 
would  help  the problem. But in any case,  no clear 
conclusions on the viability of the self-accelerated  solution can be  reached 
without considering the quantum effects, which by themselves   call for 
an appropriate computational scheme since the theory is strongly coupled.  

Once the theory of massive gravity is embedded in the framework of 
\cite {Gabadadze:2014rwa,ggsy}, the quantum loop corrections in the particle physics sector are governed
by $\hbar$, while the loops in the gravitational sector 
are governed by an effective Planck's constant, $\hbar {\tilde q} $, $\tilde q $ being a functional whose value is determined by the classical equations of motion. 
On the solutions considered in \cite {Gabadadze:2014rwa,ggsy}, one gets $\tilde q \to 0$, 
so the quantum loop corrections 
in the gravitational sector vanish. Thus, at low energies, the tensor, vector, and scalar modes  of gravity need not be 
quantized. Then the vanishing of the vector field quadratic terms
is not a problem for a self-accelerated solution with healthy scalar and tensor 
perturbations, since the solution ``vector $=0$" can be maintained perpetually 
in the absence of quantum fluctuations. 

In summary, we calculate corrections 
to the decoupling limit Lagrangian of massive gravity due to the nonlocal terms
used to solve the CC problem in \cite {Gabadadze:2014rwa,ggsy}. 
We show that the obtained theory with the new nonlocal 
Galileon terms has a self-accelerated solution with healthy scalar and tensor fluctuations  
for $a_3=0$, in which case the tensor and scalar 
nonlinear mixing terms can be diagonalized. A special feature of this background is that there is no 
linear scalar-tensor mixing to begin with, and therefore no vDVZ discontinuity \cite {dRGHP_sa}.  
Thus, in the approximation used, linear massive gravity is indistinguishable 
from linear  general relativity  on a de Sitter background of equal curvature; the  two theories  differ only  by  nonlinear  terms.   The vector field still loses its quadratic terms, 
like it does in pure massive gravity, but this is not an issue since we 
can impose ``vector $=0$" without being driven away by the quantum loops.  We also investigate spherically symmetric solutions 
on the self-accelerated background and show their stability for certain 
values of the parameter $a_2$.

Last but not least, C. Cheung and G. Remmen have recently derived the constraints on the pure massive gravity parameters imposed by unitarity and analyticity of scattering amplitudes \cite {cheung}. Since our nonlocal interactions leave local physics intact, the same constraints would apply to our theory as well. The range of parameters used for the healthy self-accelerated solutions in this work, $0<a_2<1/6$ and $a_3=0$, lies entirely in the allowed region of the parameter space 
computed in \cite {cheung}.

\section{The Framework}\label{theory}

The goal is to cancel the big CC via a stable mechanism of Refs.  \cite {Gabadadze:2014rwa, ggsy}, 
and introduce dark energy by massive gravity. Although our main discussions and results 
are equally  applicable  to both 4D models of Ref. \cite {Gabadadze:2014rwa},  and the 5D model of Ref. 
\cite {ggsy}, for concreteness, from now on we will focus  on  the framework  of Ref. \cite {ggsy}: 
the 4D physical metric, $g_{\mu \nu}$,  lives on the conformal boundary of a 
5D anti-de Sitter (AdS$_5$) space; the latter is endowed by its own 5D metric $F_{AB}$.
In other words,  a  spacetime boundary  -- or an orbifold plane -- has its own 4D metric  $g$  and   is  placed right at the  conformal boundary of AdS$_5$  
described by the metric $F$.  The metrics $g$ and $F$ need not be related to each other. 
The theory  is described by  the  following nonlocal  action functional 
\begin{equation}\label{pr1}
\mathcal{A}=\frac{V_F}{V_g}S+S_F,
\end{equation}
where $V_F$ and $V_g$ are the respective invariant volumes of the 5D  and 4D spacetimes, 
$S_F$ is the bulk Einstein-Hilbert action with a large negative CC, and $S$ is the action of our world, 
containing $g$-gravity and matter fields. 

The action (\ref{pr1}) exhibits  the following property:    
any CC generated in the 4D $g$-universe  by any source --  classical or quantum -- 
is immaterial for the  $g$-universe. Instead, that CC gets absorbed into $S_F$;  
it adds to, or subtracts from,  the curvature of AdS$_5$. 

One should expect  from  (\ref {pr1}) that the gravity and the  particle physics Lagrangians each require 
special treatment at the quantum level \cite{Gabadadze:2014rwa,ggsy}.  
The upshot  of such a treatment  is  that the particle physics quantum effects can be maintained  intact, 
while the gravity  loops  of $g$ are suppressed by the effective Planck constant, 
$\hbar V_g/V_F$. The ratio of the volumes is determined by the classical solutions such that $\hbar V_g/V_F\to 0$. Then the gravity loops do not generate 
any counterterms  to the  action  for the metric $g$, and 
at low energies, gravity in our universe becomes a purely classical theory that  couples to quantized particle 
physics; $F$-gravity is also quantized in a conventional way   \cite{Gabadadze:2014rwa,ggsy}.  

Dark energy is then introduced  via massive gravity. The action  $S$ that enters (\ref  {pr1}) is
\begin{equation}\label{a1}
S=\int d^4x~\sqrt{-g}\left[M_{\text P}^2R-\frac{M_{\text P}^2m^2}{4} \mathcal{U} \left( \mathcal{K}\right)+2\mathcal{L}_{\rm SM} \right]. 
\end{equation}
Here, the graviton mass and the nonlinear interactions are  described  by the dRGT potential \cite{mgr1,mgr2},
\begin{equation}
\mathcal{U}(\mathcal{K})=\det_2(\mathcal{K})+\alpha_3\det_3(\mathcal{K})+\alpha_4\det_4(\mathcal{K}),
\end{equation}
where  the matrix $\mathcal{K}=1-A$, $A$ being defined  as  
$A^{\mu}_{~ \alpha}A^{\alpha}_{ ~\nu}=g^{\mu \alpha}\gamma_{\alpha \nu}$, so that 
$\mathcal{K}=1-\sqrt{g^{-1}\gamma}$. The parameters $m, \alpha_3,\alpha_4$ are technically natural,
even if the value of $m$ is chosen to be tiny \cite {dRGHP} (for  reviews on theoretical and observational 
aspects of massive gravity, see \cite {Kurt,Claudia,Fred}).

The massive gravity theory built this way was shown to  be ghost-free \cite{ghostfree}. The fixed 
metric $\gamma$,  known as the fiducial metric,  is generally non-flat \cite{hrm}. 
Its origin is unspecified  in the pure dRGT massive gravity, but in (\ref{pr1}), it is directly related to the pullback of the boundary metric induced by bulk dynamics, and specified by the Dirichlet data that determines the bulk metric and sources the boundary CFT fields \cite{ggsy}. Here, we will make the simplest  choice of  trivial boundary data and a Minkowski fiducial metric, $\gamma_{\mu \nu}=\eta_{\mu \nu}$. This is consistent with the AdS/CFT prescription \cite{adscft}. Meanwhile, the matter fields are still quantized with $\hbar$, so the matter Lagrangian, $\mathcal{L}_{\rm SM}$, should be regarded as the real part of the full quantum 1PI Lagrangian coupled to classical gravity \cite{ggsy}.

Following a useful practice in massive gravity, we study the theory (\ref{a1}) by taking the decoupling limit: $M_{\rm P}\rightarrow \infty, m\rightarrow 0, \Lambda_3=(M_{\rm P}m^2)^{1/3}$ and $T_{\mu \nu} / M_{\rm P}$ held fixed.
While doing so we do not take any limit for the $F$-gravity since it's just a conventional massless theory. We calculate the new nonlocal terms and find stable vacuum self-accelerated solutions for the so-called  restricted Galileons \cite{restgal}, that correspond to a special choice of parameters  in a general Galileon theory \cite {nic}. We emphasize the difference between the self-accelerated solutions in the new theory and in pure massive gravity; in a certain parameter range, the former is ghost-free, but the latter is not \cite{dRGHP_sa}. We then study Schwarzschild solutions on these self-accelerated backgrounds, showing that  
the helicity-0 mode decouples from the source already in the linearized theory \cite {dRGHP_sa}. 

We adopt the mostly plus metric convention in this work. The Levi-Civit\`a symbol, $\varepsilon_{\mu \nu \alpha \beta}$, is normalized so that $\varepsilon_{0123}=1$, $\varepsilon^{0123}=-1$. The contractions of rank-2 tensors will be often denoted by square brackets, for example, $[\Pi]=\Pi^{\mu }_{~\mu}$, $[\Pi^2]=\Pi^{\mu }_{~\nu}\Pi^{\nu}_{~\mu}$, etc. We will also use the following shorthand notations, $\varepsilon \varepsilon \Pi \Pi=\varepsilon^{\mu \alpha \rho \sigma}\varepsilon^{\nu \beta}_{~~~\rho \sigma}\Pi_{\mu \nu}\Pi_{\alpha \beta}$, $\varepsilon_{\mu}\varepsilon_{\nu}\Pi\Pi=\varepsilon_{\mu}^{~~\gamma \alpha \rho}\varepsilon_{\nu \gamma}^{~~~\beta \sigma}\Pi_{\alpha \beta}\Pi_{\rho \sigma}$, $(B^2)^{\mu}_{~\nu}=B^{\mu}_{~\alpha} B^{\alpha}_{~\nu}$, $\varepsilon\varepsilon B\partial A=\varepsilon_{\mu_1 \mu_2 \alpha \beta}\varepsilon^{\nu_1 \nu_2 \alpha \beta}B^{\mu_1}_{~\nu_1}\partial_{\nu_2}A^{\mu_2}$, and so on. A ratio between spacetime integrals is defined in the following way: first compute the ratio of integrals in a finite spacetime volume, then take the limit when the region of integration tend to infinity in all dimensions.

\section{New Terms in the Decoupling Limit}\label{theo}

To  begin with, let us derive the  the new terms that emerge in the decoupling limit 
of massive gravity  due to the nonlocal interactions. The variation of (\ref{a1}) yields
\begin{equation}
\delta \bar{S}\equiv \delta\left( \frac{S}{V_g}\right)=\frac{1}{V_g}\left(\delta S-\frac{S}{V_g}\delta
V_g \right),
\end{equation}
where $V_g$ is a functional of the helicity-2 field, $h_{\mu \nu}$, but depends neither on the helicity-0 field, $\pi$, nor  on 
the helicity-1 field, $A_{\mu}$. Define $h_{\mu \nu}/M_{\rm P}\equiv g_{\mu \nu}-\eta_{\mu \nu}$, then
\begin{equation}\label{expand0}
V_g=\int d^4x~\left[ 1+\frac{h}{2M_{\text P}}-\frac{1}{4M_{\text P}^2}\left( h^{\mu \nu}h_{\mu \nu}-\frac{1}{2}h^2\right)+\mathcal{O}\left(\frac{1}{M_{\text P}^3} \right) \right].
\end{equation}
The expansion in (\ref{expand0}) is governed  by  inverse powers of $M_{\text P}$. 
We'd like to determine which terms in the product of $S$  and $\delta V_g$
survive the decoupling limit in addition to the  known terms  obtained 
in that limit of massive gravity   \cite{mgr1}. Since the new terms in 
(\ref{expand0}) are of the order $1/M_{\text P}$, or smaller, 
they can give finite terms in the decoupling limit  only if they multiply 
terms in $S$ that are proportional to $1/m^2$ or its higher powers. However, there are only  three such terms
in $S$ --  these  are precisely the nonlinear terms, made of second derivatives of $\pi$, that  are  
total derivatives   \cite{mgr1}.\footnote{Note that adding a  linear term  $[\mathcal{K}]$  in pure massive gravity 
is equivalent to adding, up to a total derivative, a CC to massive gravity. In the present 
framework the CC is removed; however, the total derivative, once multiplied 
by $\delta V_g$,   would give yet another  finite nonlocal term of the form $\int d^4y h \int d^4x \partial^2 \pi$,  
as well as a tadpole, $\int d^4y h \int d^4x $, with an arbitrary finite coefficient.  
Hence,  the sum of the three total derivative terms in (\ref {TD3}) should be amended by the term proportional to unity, 
and one more total derivative term proportional to $\partial^2 \pi$, with an arbitrary finite 
coefficient.  We  do not  include the $[\mathcal{K}]$   term here in order to  avoid the tadpole and 
also to make comparison with pure massive gravity.}  
It is then necessary to keep them in the full  Lagrangian as we will do below
\begin{equation}
\mathcal{L}=\mathcal{L}_{\text{MG}}+\frac{1}{m^2}\mathcal{L}^{\text{TD}}_{\text{tot}},
\label{TD}
\end{equation}
where
\begin{equation}
\mathcal{L}_{\text{MG}}=-\frac{1}{2}h^{\mu \nu}\mathcal{E}^{\alpha \beta}_{\mu \nu}h_{\alpha \beta}+h^{\mu \nu}\sum_{n=1}^3\frac{a_n}{\Lambda_3^{3n-3}}X^{(n)}_{\mu \nu}+\frac{h^{\mu \nu}T_{\mu \nu}}{M_{\rm P}},
\end{equation}
is the  decoupling limit Lagrangian of massive gravity \cite{mgr1}, and  
$\mathcal{E}^{\alpha \beta}_{\mu \nu}$ is the Einstein operator, and the $1/m^2$ terms read  as follows
\begin{equation}
\mathcal{L}^{\text{TD}}_{\text{tot}}=\mathcal{L}^{\text{TD}}_2-\frac{1}{\Lambda_3^3}\left( \frac{2a_2}{3}+\frac{1}{3}\right) \mathcal{L}^{\text{TD}}_3+\frac{1}{\Lambda_3^6}\left(\frac{a_2}{6}-\frac{a_3}{2}+\frac{1}{12}\right)\mathcal{L}^{\text{TD}}_4, 
\label{TD3}
\end{equation}
where we have used the following notations:
\begin{gather}
\Pi_{\mu \nu}=\partial_{\mu}\partial_{\nu}\pi, \quad a_1=-\frac{1}{2},\\
X^{(1)}_{\mu \nu}=\varepsilon_{\mu}\varepsilon_{\nu}\Pi,\quad
X^{(2)}_{\mu \nu}=\varepsilon_{\mu}\varepsilon_{\nu}\Pi\Pi,\quad
X^{(3)}_{\mu \nu}=\varepsilon_{\mu}\varepsilon_{\nu}\Pi\Pi\Pi, \\
\mathcal{L}^{\text{TD}}_2=-\frac{1}{2}\varepsilon\varepsilon\Pi\Pi,\quad
\mathcal{L}^{\text{TD}}_3=-\varepsilon\varepsilon\Pi\Pi\Pi,\quad
\mathcal{L}^{\text{TD}}_4=-\varepsilon\varepsilon\Pi\Pi\Pi\Pi.
\end{gather}
Then, the helicity-2 equations  with the new terms included,  will read 
\begin{equation}\label{newh}
-\mathcal{E}^{\alpha \beta}_{\mu \nu}h_{\alpha \beta}+\sum_{n=1}^3\frac{a_n}{\Lambda_3^{3n-3}}X^{(n)}_{\mu \nu}=\eta_{\mu \nu}\Pi_*-\frac{T_{\mu \nu}}{M_{\rm P}},
\end{equation}
where
\begin{equation}
\Pi_*=\frac{1}{2\Lambda^3_3}\frac{\int d^4x~\mathcal{L}^{\text{TD}}_{\text{tot}}}{V_0},\quad  V_0= \int d^4x~1.
\end{equation}
In the latter expression we also assume that the four-volume  is regularized by some infrared spacetime box, and that the regularization is removed after all the other calculations are  done. 
The term denoted by $\Pi_*$  is the  new term. It  distinguishes the above equation from its counterpart 
in pure massive gravity obtained in \cite {mgr1}.    
Note that $\Pi_*$ depends only on the asymptotic values of $\Pi_{\mu \nu}$ 
and is suppressed by $V_0$; its variation with respect to the dynamical fields is therefore zero.  
The helicity-1 field, $A_{\mu}$, decouples completely from the tensor mode 
but appears in the Lagrangian only at the quadratic and higher orders; thus, we can consistently set it to zero 
for simplicity.  Last but not least, the helicity-0 equation of motion is the same as in pure massive gravity 
\cite{mgr1},
\begin{equation}\label{newpi}
\partial_{\alpha}\partial_{\beta}h^{\mu \nu}\left( -\frac{1}{2}\varepsilon_{\mu}^{~\alpha} \varepsilon_{\nu ~}^{~\beta}+\frac{2a_2}{\Lambda_3^3}\varepsilon_{\mu}^{~\alpha  } \varepsilon_{\nu}^{~\beta }\Pi+\frac{3a_3}{\Lambda_3^6}\varepsilon_{\mu}^{~\alpha } \varepsilon_{\nu}^{~\beta }\Pi\Pi  \right)=0.
\end{equation}
The obtained equations (\ref{newh}) -- (\ref{newpi}) constitute the field equations of our theory 
with the most  general parameters $a_2, a_3$.

In what follows it will also be useful to cast the decoupling limit action into a more conventional form by using the expansion (\ref{expand0})  for $V_g^{-1}$.  The  new term emerging in the decoupling limit  in addition to the ones 
of pure massive gravity is given by the product of $\sim h/M_{\text{P}}$ with the integral of a 
total derivative.  Thus, the  decoupling limit action  can be written as 
\begin{equation}\label{lmg}
\bar{S}_{\rm DL}=\frac{1}{V_0}\int d^4x~ \bar{\mathcal{L}},\quad \bar{\mathcal{L}}= \mathcal{L}_{\text{MG}}-h\Pi_*.
\end{equation}
This could be used to  expand $\bar{\mathcal{L}}$ to study small fluctuations, just like in a local theory. A quick check shows that the variation of $\bar{S}_{\rm DL}$ with respect to the fields gives back the equations of motion, (\ref{newh}) and (\ref{newpi}), as it should be the case. Note also that the above action is invariant 
under linearized  diff-transformations, $h_{\mu\nu}\to  h_{\mu\nu} + \partial_\mu  \zeta_\nu+  \partial_\nu  \zeta_\mu$.

\section{Self-Acceleration with Nonlocal Galileons} \label{dd}

Here, we look for self-accelerated solutions in  the theory described in the last section. 
We focus on the case $a_3=0$, which makes the helicity-2 and -0
modes separable under the local field re-definition \cite {mgr1}.
Without loss of generality, fields in the self-accelerated background can be parametrized by 
\be \label{ansatz}
h_{\mu \nu}=-\frac{1}{2} M_{\rm P} H^2x^2\eta_{\mu \nu},\quad \pi=\frac{1}{2} q\Lambda^3_3x^2.
\ee
The above expression for  $h_{\mu\nu}$ guarantees that the Ricci curvature of the background
is $H^2$; the latter and the parameter $q$  are the unknowns of the system.   Furthermore, 
the helicity-2 and -0 equations reduce to, respectively,
\begin{gather}
\frac{M_{\rm P}H^2}{\Lambda_3^3}=-q+(2a_2+2)q^2-\frac{4}{3}(2a_2+1)q^3+\frac{1}{3}(2a_2+1)q^4, \label{H2} \\
2a_2q-\frac{1}{2}=0,
\end{gather}
and the solutions  are $q=\frac{1}{4a_2}$, and 
\begin{equation}\label{pi0}
H^2=\frac{(1-4a_2)^2(1-6a_2)}{768a_2^4}m^2.
\end{equation}
The  latter expression for $H^2$  is positive only for $a_2<1/6$.

Note that in the conventional decoupling limit, $m\to 0$, and therefore $H\to 0$  if the 
parameter $a_2$ is held fixed. Thus the limit corresponds to the physical  approximation that all the distance/time scales considered are much smaller than the 
Hubble distance/time scale set by $H^{-1}$.

To check the linear stability of the above obtained solutions, we diagonalize the theory as in  \cite {mgr1}, 
whereby the pure massive gravity Lagrangian takes the form $\mathcal{L}_{\rm MG}=\mathcal{L}_{\bar{h}}+\mathcal{L}_{\pi}$, $\mathcal{L}_{\bar{h}}$ being the linearized Einstein-Hilbert terms minimally coupled to $T_{\mu \nu}$, and
\begin{equation}\label{old}
\mathcal{L}_{\pi}=\frac{3}{2}\pi \square \pi+\frac{3a_2}{\Lambda_3^3}(\partial\pi)^2\square\pi+\frac{2a_2^2}{\Lambda_3^6}(\partial\pi)^2\left([\Pi^2]-[\Pi]^2\right)+\frac{\pi T}{M_{\rm P}}+\frac{2a_2}{M_{\rm P}\Lambda_3^3}\partial_{\mu}\pi\partial_{\nu}\pi T^{\mu \nu}.
\end{equation} 
This is similar to the cubic and quartic Galileon theories with special coefficients.\footnote{Thus, the diagonalizable subclass of theories are sometimes called ``restricted Galileons," and the non-diagonalizable ones with $a_3\neq 0$, ``mixed Galileons" \cite{restgal, Berezhiani:2013dca}.} Clearly, the helicity-2 modes are automatically stable and exactly luminal. However, for the helicity-0 field, the new nonlocal  term contributes as well; the total 
effective Lagrangian for $\pi$ is
\begin{equation}
\bar{\mathcal{L}}_{\pi}=\mathcal{L}_{\pi}-\Pi_*\left[ \frac{2a_2}{\Lambda_3^3}(\partial{\pi})^2+4\pi \right].
\end{equation}
While the $\pi$ tadpole here is not itself invariant under the Galilean transformation, its variation is cancelled by that of $\bar{h}_{\mu 
\nu}$, since $h_{\mu \nu}$ as a whole does not transform. Now, if $a_2<0$, then the last term in (\ref{old}) would have a negative classical renormalization of the $\pi$ kinetic term. For most of the reasonable sources, this would overshoot the sign of $\pi$ kinetic term, whereas the nonlinear terms in (\ref{old}) will only be subdominant \cite{restgal}. Hence, the physical choice for the sign of $a_2$ is $a_2>0$. Combining this with the positivity condition for $H^2$ above, we get $0< a_2 < 1/6$. We will see below that $a_2 < 1/6$ is also necessary to avoid ghostly fluctuations.

Next, consider small perturbations about  a background solution $\pi_b$, 
$\pi=\pi_b+\phi$. Then the Lagrangian,  up to the second order in $\phi$, reads as 
\begin{equation}\label{newphi}
\bar{\mathcal{L}_{\phi}}=\mathcal{L}_{\phi}-\Pi_* \left[ \frac{2a_2}{\Lambda_3^3}(\partial{\phi})^2 +4 \phi \left( 1- \frac{a_2}{\Lambda_3^3}\square \pi_b \right) \right],
\end{equation}
where
\begin{equation}\label{oldmgr}
\begin{aligned}
\mathcal{L}_{\phi}=&\left\{ \left[  -\frac{3}{2}+\frac{6a_2}{\Lambda_3^3}\square \pi_b+\frac{6a_2^2}{\Lambda_3^6}\left((\partial_{\alpha}\partial_{\beta}\pi_b)^2-( \square \pi_b)^2 \right) \right]\eta_{\mu \nu} \right. \\
& \left.-\frac{6a_2}{\Lambda_3^3}\partial_{\mu}\partial_{\nu}\pi_b+\frac{12a_2^2}{\Lambda_3^6}(-\partial_{\alpha}\partial_{\mu}\pi_b\partial^{\alpha}\partial_{\nu}\pi_b+\square\pi_b\partial_{\mu}\partial_{\nu}\pi_b)\right\}\partial^{\mu}\phi\partial^{\nu}\phi  \\
&+\frac{\phi}{M_{\rm P}}\left( T-\frac{4a_2}{\Lambda_3^3}\partial_{\mu}\partial_{\nu}\pi_b T^{\mu \nu} \right) +\frac{2a_2}{M_{\rm P}\Lambda_3^3}\partial_{\mu}\phi\partial_{\nu}\phi T^{\mu \nu}\,,
\end{aligned}
\end{equation}
contains all the terms obtained in pure massive gravity \cite{restgal}. 

Plugging in the solutions (\ref{ansatz}) -- (\ref{pi0}), the quadratic Lagrangian for $\phi$ becomes
\begin{equation}\label{newperturblag}
\bar{\mathcal{L}}_{\phi}=-\frac{(1-4a_2)^2(1-6a_2)}{128a_2^3}(\partial\phi)^2  +\frac{2a_2}{M_{\text P}\Lambda_3^3}\partial_{\mu}\phi\partial_{\nu}\phi T^{\mu \nu}.
\end{equation}
Note that the $\phi$ tadpole term vanishes, confirming the kinetic de-mixing between the helicity-2 and -0 fluctuations on self-accelerated backgrounds. This happens because the coefficient of the kinetic mixing term is proportional to the $\pi$ equation of motion  \cite{dRGHP_sa}, and therefore the de-mixing is not affected by any cosmological matter present in  $T_{\mu \nu}$. The expression  (\ref{newperturblag}) shows that the scalar fluctuations on the self-accelerated backgrounds with $0<a_2<1/6$ are safely stable and non-superluminal. This differs significantly  from 
pure massive gravity. Indeed, without the extra contribution from the second term in (\ref{newphi}), the quadratic Lagrangian would have been generated by (\ref{oldmgr}) only  and would have taken the form,
\begin{equation}
\mathcal{L}_{\phi}=\frac{3}{4}(\partial\phi)^2+\frac{2a_2}{M_{\text P}\Lambda_3^3}\partial_{\mu}\phi\partial_{\nu}\phi T^{\mu \nu}.
\end{equation}
The latter describes  a ghost for all self-accelerated solutions, as already seen in the literature \cite{dRGHP_sa}. 
Pure massive gravity requires $a_3\neq 0$ for the scalar perturbation to have a positive kinetic term; 
in the present theory  one can get  away even with $a_3=0$, as described above.

Note that should $a_2>1/6$ (and $\neq 1/4$), the resulting AdS solution with $H^2<0$ 
would have carried a ghost. 

\section{Spherically Symmetric Source on Self-Accelerated Background} \label{3.3} 

In this section, we study the gravitational field configuration created by a spherically symmetric 
source of finite size $R$ and uniform density and pressure, using the techniques of Ref. \cite {GigaDavid}; 
the source has total mass $M$. We would like to see if a  Schwarzschild-like solution exists on the 
healthy self-accelerated backgrounds found above.

We start with the most general spherically symmetric metric parametrization,
\begin{gather}
h_{00}=a(r),\quad h_{ij}=f(r)\delta_{ij}, \\
\pi=\frac{c}{2}\Lambda_3^3t^2+\pi_0(r).
\end{gather}
By integrating the helicity-2 equations, (\ref{newh}), with the condition  that  the fields  
vanish at the origin, we'll   get 
\begin{gather}
\frac{f'}{r}=-\frac{\Pi_*}{3}-\frac{M(r)}{4\pi M_{\text P}r^3}+\Lambda_3^3(\lambda-2a_2\lambda^2),  \label{he21}  \\
\frac{a'}{r}=\frac{2\Pi_*}{3}-\frac{M(r)}{4\pi M_{\text P}r^3}+\Lambda_3^3 \lambda(1+4ca_2)+c\Lambda_3^3. \label{he22}
\end{gather}
Then,  the helicity-0 equation, (\ref{newpi}),  reduces to the following expression
\begin{equation}\label{mastereq}
\begin{aligned}
&8a_2^2\lambda^3-12a_2(1+2ca_2)\lambda^2+\left[ 3+4a_2\left( 3c+\frac{\Pi_*}{\Lambda_3^3} \right) \right]\lambda -\frac{4\Pi_*}{3\Lambda_3^3}(1+ca_2)-c  \\
&=\left\{ \begin{aligned}
&\left( \frac{r_*}{r} \right)^3 (1+4ca_2), \quad r>R, \\
&\left( \frac{r_*}{R} \right)^3(1+4ca_2), \quad r<R. \\
\end{aligned}
\right.
\end{aligned}
\end{equation}
Above  we have used the notations  
\begin{equation}
\lambda(r)=\frac{\pi_0'(r)}{\Lambda_3^3r}, \quad r_*=\left(\frac{M}{4\pi M_{\text P}^2m^2} \right)^{\frac{1}{3}},
\end{equation}
where $r_*$ is known as the Vainshtein radius. Let us calculate the  expression for the 
nonlocal term on the ansatz at hand,
\begin{equation}\label{pistar}
\Pi_*=\frac{1}{2\Lambda^3_3}\frac{\int d^4x~\mathcal{L}^{\text{TD}}_{\text{tot}}}{V_0} =\Lambda^3_3[-3c\lambda_{\infty}-(1+c)(1+2a_2)\lambda_{\infty}^3+3\lambda_{\infty}(1+c+2ca_2)],
\end{equation}
where $\lambda_{\infty}\equiv \lambda(r \rightarrow \infty)$.

We can see that there exists a ``de-mixing" value,  $c=-1/(4a_2)$, which makes the helicity-0 equation independent of the mass of the source. This defines a class of solutions on which the kinetic mixing between helicity-2 and -0 fluctuations vanishes. In this case, $\lambda$ must be constant on solution and equal to its asymptotic value, $\lambda_{\infty}$. The expression (\ref{mastereq}) then becomes an equation for $\lambda_{\infty}$,
\begin{equation}\label{demixroot}
(\lambda_{\infty}-1)(4a_2\lambda_{\infty}-1)[(8a_2^2+2a_2-1)\lambda_{\infty}^2+(2-4a_2)\lambda_{\infty}-1]=0.
\end{equation}
In particular, the solutions with $\lambda_{\infty}=1/(4a_2)$ corresponds to the self-accelerated background found in the last section.\footnote{The other roots of (\ref{demixroot}) also give valid Schwarzschild solutions, only that they correspond to pressureless, positive-energy cosmological backgrounds. We discuss them in Appendix \ref{cosb}.} 

 Solving for the metric components, we obtain
\begin{equation}
a=\frac{M}{4\pi M_{\text P}r}+r^2\frac{(1-4a_2)^2(1-6a_2)}{768a_2^4}\Lambda_3^3,
\end{equation}
\begin{equation}
f=\frac{M}{4\pi M_{\text P}r}-r^2\frac{(1-4a_2)^2(1-6a_2)}{1536a_2^4}\Lambda_3^3.
\end{equation}
Therefore, the effective pressure and energy density of the $\pi$ fluid are, respectively,\footnote{Note that $p$ and $\rho$ here are not calculated from the canonically defined stress-energy tensor, $T_{\mu \nu}$, but from the linearized Einstein tensor. In view of (\ref{newh}), they receive extra contributions from the Galileon and the nonlocal terms. This comment applies to all pressures and energy densities calculated in this paper.}
\begin{equation}
p=\frac{M_{\text P}}{3}\nabla^2(f-a)=-\frac{(1-4a_2)^2(1-6a_2)}{256a_2^4}M_{\text P}\Lambda_3^3,
\end{equation}
\begin{equation}
\rho=-M_{\text P}\nabla^2f=\frac{(1-4a_2)^2(1-6a_2)}{256a_2^4}M_{\text P}\Lambda_3^3,
\end{equation}
with the right dS equation of state $p=-\rho$. The curvature can then be restored,
\begin{equation}
H^2=\frac{\rho}{3M_{\rm P}^2}=\frac{(1-4a_2)^2(1-6a_2)}{768a_2^4}m^2,
\end{equation}
and coincides, as it should,  with  that obtained in a different gauge in Section \ref{dd}. 
Owing to the kinetic de-mixing, the stability analysis is identical to that of the source-free 
case, so the obtained  Schwarzschild solutions  are stable in the range, $0<a_2<1/6$.

So far, we have only considered a special value of the parameter $c$  that leads to the de-mixing phenomenon. 
We want to see if there exist other self-accelerated backgrounds on which star-like objects can survive. 
A necessary condition for such solutions is that as $r \rightarrow \infty$, the $\pi$ field profile tends to $\pi \rightarrow c\Lambda_3^3(t^2-r^2)/2$; this  implies that $c=-\lambda_{\infty}$. By requiring this in (\ref{mastereq}), we obtain
the following algebraic equation 
\begin{equation}\label{selfa}
4\lambda_{\infty}-8\lambda_{\infty}^2(1+3a_2)+\frac{16}{3}\lambda_{\infty}^3(1+8a_2+6a_2^2)-\frac{4}{3}\lambda_{\infty}^4(1+18a_2+32a_2^2)+\frac{16}{3}\lambda_{\infty}^5a_2(1+2a_2)=0,
\end{equation} 
and look for its roots. These roots are 
\begin{equation}\label{roots}
\lambda_{0,1,2,3,4}=\frac{1}{4a_2},~0,~1,~\frac{3}{2}-\frac{\sqrt{3}}{2}\sqrt{\frac{6a_2-1}{2a_2+1}},~\frac{3}{2}+\frac{\sqrt{3}}{2}\sqrt{\frac{6a_2-1}{2a_2+1}}.
\end{equation}
Among them, $\lambda_{\infty}=\lambda_{0}$ corresponds to the de-mixing case already discussed. As to  
the rest of the roots, the helicity-0 equation does depend on the mass of the source, so if $r_*>R$, there is a Vainshtein region sufficiently close to the course, in which $\lambda \gg 1$. This must be able to extrapolate to $\sim \lambda_{\infty}$ outside the Vainshtein region, so $\lambda_{\infty}$ has to be the largest root of (\ref{mastereq}) at $r\rightarrow \infty$, with $c$ set to $-\lambda_{\infty}$ \cite{nic}. Given that $a_2>0$, this can never be satisfied when $\lambda_{\infty}=\lambda_{1,3}$, for which there is always a larger root. In particular, the existence of the 
obstruction   for  $\lambda_{\infty}=0$  implies that asymptotic flatness must be given up in the presence of a spherically symmetric source. The other cases, $\lambda_{2,4}$, are able to satisfy the interpolation condition given appropriate values of $a_2$, resulting in stable cosmological solutions that are, however, not self-accelerated. We give explicit forms and stability analyses of them in Appendix \ref{cosb}. To conclude, the dS vacua given by $\lambda_{\infty}=1/(4a_2)$, $0<a_2<1/6$ are the unique,  stable, non-superluminal,  self-accelerated backgrounds that have Schwarzschild solutions exhibiting the de-mixing phenomenon.  

\section*{Acknowledgements}

Both GG  and SY  were   supported by  NASA  grant  NNX12AF86G  S06 and 
by NSF  grant PHY-1316452. GG acknowledges a membership at 
the NYU-ECNU Joint Physics Research Institute in Shanghai.

\appendix

\section{Schwarzschild Solutions on Cosmological Backgrounds}\label{cosb}

\subsection{De-mixing Solutions}
The choice $c=-1/(4a_2)$ leads to kinetic de-mixing between the helicity-2 and -0 modes. To find more de-mixing solutions, the other roots of (\ref{demixroot}) are
\begin{equation}
\lambda_{\infty}=1,~\frac{2a_2-1\pm\sqrt{2a_2(6a_2-1)}}{8a_2^2+2a_2-1}.
\end{equation}
We look at them case by case.

$\lambda_{\infty}=1$: We have 
\begin{equation}
a=f=\frac{M}{4\pi M_{\text P}r}-r^2\frac{(1-4a_2)^2}{24a_2}\Lambda_3^3,\quad p=0, \quad \rho=\frac{(1-4a_2)^2}{4a_2}M_{\text P}\Lambda_3^3,
\end{equation}
\begin{equation}
\bar{\mathcal{L}}_{\phi}=2(1-4a_2)^2\left[(\partial_t\phi)^2-\frac{1}{4}(\partial_j\phi)^2 \right],
\end{equation}
so the scalar fluctuation is stable, propagates at half the speed of light, and has positive energy density if $a_2>0$ and $a_2\neq 1/4$.

$\lambda_{\infty}=\lambda_{\pm}=\frac{2a_2-1\pm\sqrt{2a_2(6a_2-1)}}{8a_2^2+2a_2-1}$: We have
\begin{equation}
a=f=\frac{M}{4\pi M_{\text P}r}-r^2\frac{F_{\mp}}{24a_2}\Lambda_3^3,\quad p=0, \quad \rho=\frac{F_{\mp}}{4a_2}M_{\text P}\Lambda_3^3,
\end{equation}
\begin{equation}
\bar{\mathcal{L}}_{\phi}=2F_{\mp}\left[(\partial_t\phi)^2-\frac{1}{4}(\partial_j\phi)^2 \right],
\end{equation}
where
\begin{equation}
F_{\pm}=\frac{(6a_2-1)(32a_2^3+6a_2-1\pm 8a_2\sqrt{2a_2(6a_2-1)})}{(8a_2^2+2a_2-1)^2} >0.
\end{equation}
Again, the fluctuation is stable, propagates at half the speed of light, and has positive energy density if $a_2>1/6$ and $a_2\neq 1/4$. These solutions have zero pressure and positive energy density for $\pi$ fluid.

\subsection{Solutions with Nontrivial Vainshtein Region}
With $c=-\lambda_{\infty}$, we look at the possibilities corresponding to $\lambda_{\infty}=\lambda_2,\lambda_4$ in (\ref{roots}),
\begin{equation}
\lambda_2=1,\quad \lambda_4=\frac{3}{2}+\frac{\sqrt{3}}{2}\sqrt{\frac{6a_2-1}{2a_2+1}}.
\end{equation}

$\lambda_{\infty}=\lambda_2$: One needs $a_2> 1/4$ for $\lambda_{\infty}$ to be the largest root. To leading order,\\
Outside the Vainshtein radius,
\begin{equation}
\lambda \simeq \lambda_2,\quad f \simeq \frac{M}{4\pi M_{\text P}r},\quad a \simeq \frac{M}{4\pi M_{\text P}r}+r^2\Lambda_3^3(1-4a_2),
\end{equation}
\begin{equation}
\rho \simeq 0, \quad p \simeq 2(4a_2-1)M_{\text P}\Lambda_3^3>0,
\end{equation}
\begin{equation}
\bar{\mathcal{L}}_{\phi} \simeq -\frac{3}{2}(1-4a_2)^2\partial_{\mu}\phi \partial^{\mu}\phi;  
\end{equation}
Inside the Vainshtein radius,
\begin{equation}
\lambda \simeq \frac{(1-4a_2)^{1/3}}{2a_2^{2/3}} \frac{r_*}{r}+\frac{1-2a_2}{2a_2}+\left[\frac{(1-4a_2)^{2/3}}{4a_2^{4/3}}-\frac{(1-4a_2)^{2/3}}{a_2^{1/3}} \right] \frac{r}{r_*},
\end{equation}
\begin{equation}
a\simeq f \simeq \frac{M}{4\pi M_{\text P}r}+\mathcal{O}(r^2\lambda).
\end{equation}
\begin{align}
\bar{\mathcal{L}}_{\phi} \simeq ~& 3\left[(1-4a_2)^{2/3}a_2^{2/3} \left(\frac{r_*}{r}\right)^2+2(1+(1-4a_2)^{1/3}a_2^{1/3})(1-4a_2)\frac{r_*}{r}\right](\partial_t\phi)^2  \nonumber \\
&-3\left[(1-4a_2)^{2/3}a_2^{2/3} \left(\frac{r_*}{r}\right)^2+(1-4a_2)\frac{r_*}{r}\right](\partial_r\phi)^2 \nonumber \\
&-3 \left[  \frac{1-4a_2}{2}\frac{r_*}{r}-\frac{1}{2}+\frac{(1-4a_2)^{1/3}}{4a_2^{1/3}}- (1-4a_2)^{1/3}a_2^{1/3}+8a_2^2 \right] (\partial_{\Omega}\phi)^2.  
\end{align}

$\lambda_{\infty}=\lambda_4$: One needs $a_2> 1/6$ for $\lambda_{\infty}$ to be the largest root. For notational convenience, define
\begin{equation}
\diamondsuit=36a_2^2+12a_2-3, \quad \heartsuit=\frac{12a_2^2-1+2a_2(2+\sqrt{\diamondsuit})}{1+2a_2}.
\end{equation}
Outside the Vainshtein radius,
\begin{equation}
\lambda \simeq \lambda_4,\quad f \simeq \frac{M}{4\pi M_{\text P}r},\quad a \simeq \frac{M}{4\pi M_{\text P}r}-r^2\Lambda_3^3 \frac{72a_2^2+6a_2-3+(12a_2-1)\sqrt{\diamondsuit}}{2(1+2a_2)}  ,
\end{equation}
\begin{equation}
\rho \simeq 0, \quad p \simeq \frac{72a_2^2+6a_2-3+(12a_2-1)\sqrt{\diamondsuit}}{1+2a_2}  M_{\text P}\Lambda_3^3>0,
\end{equation}
\begin{equation}
\bar{\mathcal{L}}_{\phi} \simeq -\frac{3(6a_2-1)[24a_2^2-1+4a_2(1+\sqrt{\diamondsuit})]}{2(1+2a_2)}\partial_{\mu}\phi \partial^{\mu}\phi;  
\end{equation}
Inside the Vainshtein radius,
\begin{equation}
\lambda \simeq -\frac{1}{2a_2^{2/3}}\heartsuit^{1/3} \frac{r_*}{r}-\frac{6a_2^2-1+a_2(1+\sqrt{\diamondsuit})}{2a_2(1+2a_2)}-\frac{1}{4a_2^{4/3}}\heartsuit^{5/3} \frac{r}{r_*},
\end{equation}
\begin{equation}
a\simeq f \simeq \frac{M}{4\pi M_{\text P}r}+\mathcal{O}(r^2\lambda).
\end{equation}
\begin{align}
\bar{\mathcal{L}}_{\phi} \simeq ~& 3\left[a_2^{2/3}\heartsuit^{2/3} \left(\frac{r_*}{r}\right)^2+2a_2^{1/3}\heartsuit^{4/3}\frac{r_*}{r}\right](\partial_t\phi)^2  \nonumber \\
&-3\left[a_2^{2/3} \heartsuit^{2/3}\left(\frac{r_*}{r}\right)^2 +\frac{144a_2^3-10a_2+1+4a_2(6a_2-1)\sqrt{\diamondsuit}}{2(1+2a_2)} \right](\partial_r\phi)^2 \nonumber \\
&-3\left[ \frac{144a_2^3-10a_2+1+4a_2(6a_2-1)\sqrt{\diamondsuit}}{2(1+2a_2)} \right] (\partial_{\Omega}\phi)^2.  
\end{align}

These solutions are free of ghost, gradient instability or superluminality. Inside the Vainshtein region, the radial fluctuation is slightly subluminal, and the angular one is heavily suppressed. However, far away from the source, the $\pi$ fluid has zero energy density and positive pressure, which cannot possibly match the current observational data.

\end{document}